\documentclass[preprint,showpacs,preprintnumbers,amsmath,amssymb,floats]{revtex4}
\usepackage[dvips]{graphicx}
\usepackage[english]{babel}
\usepackage{amsmath}
\usepackage{amssymb}
\usepackage{color}

\begin{document}
\title{Superconducting Transition Temperatures for Spin-Fluctuation Promoted Superconductivity in Heavy Fermion  Compounds}
 \author  {Shinya Nishiyama$^1$, K. Miyake$^1$ and C. M. Varma$^2$}
\affiliation{$^1$Department of Materials Engineering Science, Osaka University, Osaka, Japan\\ $^2$Department of Physics, University of California, Riverside, CA.}
\date{\today}
\begin{abstract}
      The quantum critical Antiferromagnetic (AFM) fluctuation spectra measured by inelastic neutron scattering  recently in two heavy fermion superconductors are used together with their other measured properties to calculate their D-wave superconducting transition temperatures $T_{\rm c}$. To this end, the linearized Eliashberg equations for D-wave superconductivity induced by AFM fluctuations are solved in models of fermions with various levels of nesting. The results for the ratio of $T_{\rm c}$ to the characteristic spin-fluctuation energy are well parametrized by a dimensionless coupling constant and the AFM correlation length. Comparing the results with experiments suggests that one may reasonably conclude that superconductivity in these compounds is indeed caused by AFM fluctuations. This conclusion is strengthened by a calculation with the same parameters of the measured coefficient of the normal state quantum-critical resistivity $\propto T^{3/2}$ characteristic of {\it gaussian} AFM quantum-critical fluctuations. The calculations give details of the superconducting coupling as a function of the correlation length and the integrated fluctuation spectra useful in other compounds. 
      \end{abstract}
\maketitle
\section{Introduction}

Many years ago, superconductivity was discovered in heavy-fermion compounds \cite{steglich, ott, stewart}. It was suggested \cite{cmv1} that the superconductivity was due to collective electronic fluctuations and not due to electron-phonon interactions.  Transport properties in the superconducting state were analyzed \cite{ssr, woelfle} to show that superconductivity was in the D-wave symmetry. It was also suggested that the
D-wave symmetry is promoted by Antiferromagnetic fluctuations \cite{miyake} with long enough correlation lengths. This promotes scattering of fermions near the fermi-surface predominantly through angles around $\pm \pi/2$, which is essential for superconducting instability in the "D-wave" channel for a suitable fermi-surface \cite{cmv-review}. The idea of long enough AFM correlation lengths as essential for this mechanism is supported by the fact that in heavy-fermions, superconductivity occurs generally in the regime near the AFM quantum critical point where the correlation lengths are long but the competing AFM phase has lower condensation energy.

 At the same time, Random phase approximation on the Hubbard model was used to calculate the spin-fluctuation spectra and to suggest that D-wave superconductivity is promoted by such fluctuations \cite{scalapino}. The properties of the Hubbard model have proven controversial in more elaborate calculations; there are calculations which suggest that the ratio of the transition temperature $T_{\rm c}$ to the typical electronic kinetic energy parameters $t$ is more than $O(10^{-2})$ \cite{DMFT} to less than $O(10^{-3})$ \cite{imada}. Since heavy-fermion properties require Kondo effect of the f-orbital local moments and their magnetic interactions using the wide-band electrons, a multi-orbital model is obviously required \cite {varma-yafet}. The Hubbard model was proposed as a sufficient model for the cuprate compounds \cite{anderson}. But the discovery in under-doped cuprates \cite{bourges} of the predicted time-reversal breaking order parameter \cite{cmv2} on the basis of a multi-orbital model raises doubts on the validity of the Hubbard model for the cuprates. For pnictides, generalization of the Hubbard model to multi-orbital situations and inclusions of Hund's rule couplings appears essential. 
 
We have a more modest goal in this paper than calculating spin-fluctuations from microscopic theory and using it to calculate properties of the superconductor.  In recent years inelastic neutron scattering in the heavy fermion compounds CeCu$_2$Si$_2$ \cite{stockert1, stockert2, arndt} and CeIrIn$_5$ \cite{kambe} have provided details of the AFM fluctuation spectrum in the normal state. The primary aim of this paper is to estimate the superconducting transition temperature using the parameters provided by the experiments in these compounds. To do so, we solve the Eliashberg equations for d-wave superconductivity using a phenomenological AFM spectral function with which the experimental data is in good accord. The use of the Eliashberg equations for quantitative calculations may be open to question because the Migdal expansion parameter, which is of $O(10^{-2})$ for the electron-phonon problem is of $O(1)$ for such compounds if one assumes that the scale of the AFM fluctuations extends to the order of the electronic bandwidth. However, when the AFM correlation length $\xi$ is large compared to the lattice constant $a$ or $(2k_{\rm F})^{-1}$, the scale of the AFM fluctuations is reduced correspondingly to $O((a/\xi)^2) t$. But in the limit of large correlation lengths, new questions arise \cite{cmv-review} which are not important in the electron-phonon problem. The most prominent among them are the role of inelastic scattering in depressing $T_{\rm c}$ on the one hand \cite{MSV}, and the fact that the BCS type coupling constant $\lambda$ appears to diverge when the characteristic fluctuation frequency $\to 0$ and the BCS prefactor appears to go to $0$.  An answer to these questions and various considerations which determine $T_{\rm c}$ from AFM interactions is possible from the numerical solution of the Eliashberg equations. 

We find that it is reasonable to conclude from a comparison of the calculated $T_{\rm c}$ with experiments that AFM fluctuations are responsible for D-wave superconductivity in the heavy fermion compounds. Very importantly, with similar parameters we calculate the {\it measured} coefficient of the anomalous $\propto T^{3/2}$ contribution to the resistivity in these compounds. A claim to quantitative accuracy on both these quantities can however be made only to factors of $O(2)$.

We note here that if one adopts that the dimensionless measure $T_{\rm c}/E_{\rm F}$ for how high is the electronic fluctuation induced superconducting, the heavy fermions may be said to do very well indeed. For example, in many cases, including the compounds studies here, this ratio is $O(10^{-2})$, similar to that of the cuprates.

Following the proposals that AFM fluctuations may also promote superconductivity in the cuprate compounds \cite{various}, there have been many discussions of the mechanism and many calculations based on the Eliashberg equations. A partial list includes the following \cite{references}. The most complete of these calculations appear to us to be those carried out by Monthoux and Lonzarich (ML) \cite{ml1,ml2}, both for 2 and 3 dimensional models. We present below calculations for the 2 dimensional square lattice model with a phenomenological spin-fluctuation spectrum, whose results are no different from those of ML for the range of parameters examined that are common. A difference in the calculations is that we vary the parameters in the two dimensional model so that "nesting" at the AFM wave-vector quantitatively changes. The amount of nesting does have a significant effect on the results. More important is that now that  the AFM fluctuation spectra is available, we can use the experimental parameters  to test the ideas quantitatively.  We also discuss how to put limits on the parameters used based on sum-rule for the fluctuation spectra and show that they are inter-related. Results for the range of physical parameters that we find relevant for the heavy fermions is not available in the published results of ML. This  has  bearing also on general conditions to determine the extent to which AFM fluctuations give significant $T_{\rm c}$ for relevant parameters in other compounds.  

This paper is organized as follows: We present in Sec. (II) the models for fermi-surface and for the spin-fluctuations which we have investigated using the linearized Eliashberg equations. We discuss there the change of effective coupling constants  with the AFM correlation length using sum-rules so that the results for  numerical solutions of the Eliashberg equations presented later are presaged.   We present the results of the calculations in Sec. (III) and discuss the important conclusions immediately after the description of the Models. We also present, in an Appendix, the explicit derivation of the coefficient of the $T^{3/2}$ resistivity from the measured form of the AFM critical fluctuation spectra. This is used in the text to estimate independently the value of a coefficient $\lambda$, which is important for the calculation of $T_{\rm c}$. We give the parameters that have been deduced by inelastic neutron scattering for the heavy fermion compounds CeCu$_2$Si$_2$ \cite{stockert1, stockert2, arndt} and CeIrIn$_5$ \cite{kambe} and compare the measured $T_{\rm c}$ with the calculations. We should emphasize that such a comparison is meant to be only illustrative of the physical principles involved; no detailed quantitative agreement is to be expected, especially given that the electronic structure of these compounds is far more complicated than assumed in the models studied. However, enough details can be provided so that one can conclude that the idea of AFM fluctuations near the quantum critical point in these compounds as the source of D-wave superconductivity is well supported. For example using measured properties, different levels of assumed nesting in the band-structure need  a coupling constant $\lambda$ between 1.5 and 3 to get the  measured $T_{\rm c}$. In this range of $\lambda$ and for the measured AFM correlation length $T_{\rm c}$ is close to being linear in $\lambda$. This range of values is compared with the value of $\lambda \approx 1.6$ needed to get the measured coefficient of the $T^{3/2}$ resistivity, which is relatively insensitive to nesting. One can assess the results from the fact that in the range of $\lambda$ deduced, $T_c$ is found to be approximately linear in $\lambda$.

\begin{figure}[tb]
	\begin{center}
		\includegraphics[width = 0.45\textwidth]{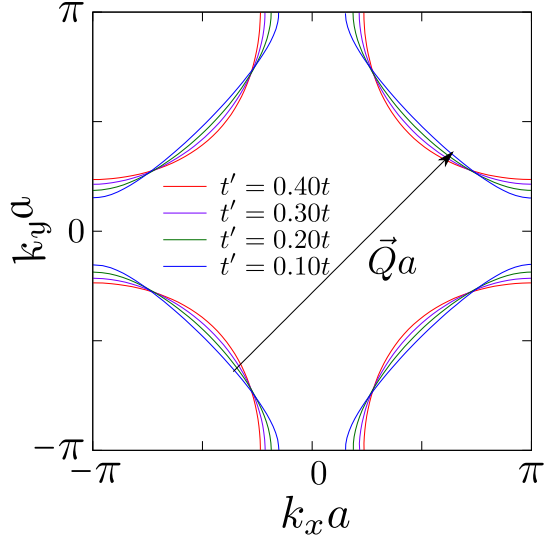}
	\end{center}
      \caption{The fermi-surfaces given by the tight binding spectrum with four values of the next nearest hopping $t'$ with filling=1.05 and t=0.3meV. }
	\label{fig:01}
\end{figure}

\section{Models and Results for $T_{\rm c}$}
\subsection{Fermi-surface}
\begin{figure}[tb]
	\begin{center}
		\includegraphics[width = 0.45\textwidth]{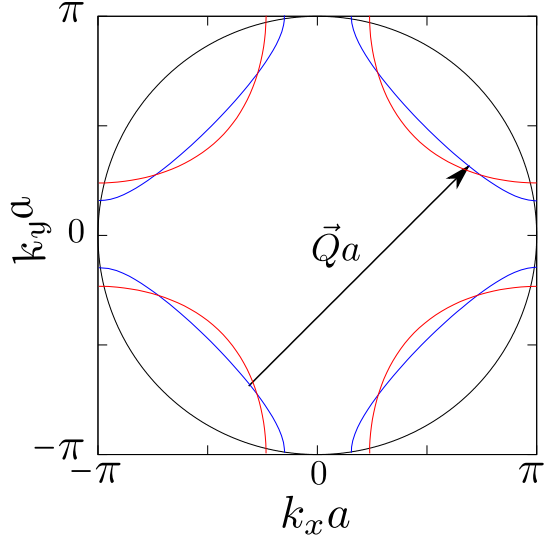}
	\end{center}
	\caption{Three types of Fermi surfaces and $\vec{Q}$-vector.
      The blue line shows the circular fermi-surface, the black line shows FS1 which is given by the tight binding spectrum in two dimensional square lattice with filling=1.05 and $t'=0.40t$.
      The red line shows FS2 which is also given by the tight binding spectrum in two dimensional square lattice but with filling=1.05, and $t'=0.10t$.}
	\label{fig:02}
\end{figure}

In our calculations we will consider two types of fermi-surfaces, a free electron fermi-surface and the others given by the tight binding spectrum in a two dimensional square lattice with nearest neighbor and next nearest neighbor hopping $t$ and $t'$ respectively:
\begin{equation}
      \varepsilon_{\vec{k}} = -2t( \cos k_x a + \cos k_y a) + 4 t^{'} \cos(k_x a) \cos(k_y a)
      \label{dis}
\end{equation}
The fermi-surface with the tight binding spectrum are shown in Fig. \ref{fig:01} for four values of the next nearest hopping $t'/t$ and the AFM wave-vector.
The nesting in the model changes as $t'$ increases.
We will show detailed result for three fermi-surfaces, the free electron fermi-surface, the fermi surface (FS1) with tight binding spectrum with $t'=0.4t$ and the fermi surface (FS2) with tight binding spectrum with $t'=0.1t$. Of the four Fermi-surfaces shown in Fig. \ref{fig:01}, the one with $t'=0.4t$ has the worst nesting and the one with $t'=0.1t$ has the best nesting. 
Fig. \ref{fig:02} shows the circular fermi-surface, FS1, FS2 and the corresponding AFM wave-vectors. 

We will discuss using the results of ML together with ours, that if properly normalized density of states and fluctuation spectra are used, two dimensional and three dimensional models give similar results for $T_{\rm c}$ provided one adjusts the ratio of the region of fermi-surface nesting to the total fermi-surface. This is in general is always lower in three than in two dimensions. It is also important to note, as discovered long ago \cite{mcmillan}, \cite{Allen-Dynes} for the case of s-wave superconductors that $T_c$ is a rather gross quantity which depends to a very good approximation on the average density of states near the chemical potential only and not on details such as the number of fermi-surface sheets and shapes. For d-wave superconductors, we we show below, it is important to also include effects of nesting of the fermi-surface near the AFM wave-vectors.

\subsection{Dynamical Spin Susceptibility for AFM Fluctuations, Correlation lengths, \\ Partial Sum rules and Coupling Constants}

The dynamical spin-susceptibility for itinerant fermions with AFM correlations may be usefully divided into two parts: The normal contribution of non-interacting fermions $\chi_0(\vec{q},i\omega)$ and the part $\chi_{\rm AFM}(\vec{q},i\omega)$ affected by AFM correlations. The two together obey the total magnetic moment sum-rule. The non-interacting susceptibility can play only an insignificant role in promoting superconductivity \cite{cmv-review} and should be ignored. The two contributions to the susceptibility can be distinguished by their momentum dependence. The characteristic momentum dependence of the non-interacting spin fluctuations is on a scale of $O(2k_f^{-1})$ while that of the AFM spin-fluctuations is much shorter. A partial sum-rule on $\chi_{\rm AFM}(\vec{q},i\omega)$ in terms of the ordered moment in the AFM phase can be used to relate the integrated fluctuations to the AFM correlation length. Double counting by using the sum-rule on the total susceptibility for the fluctuations and yet having (free) fermions interacting with such spin-fluctuations is incorrect as it over-counts the total degrees of freedom. Such considerations have usually been blithely ignored in most of the previous phenomenological work on this problem.

The fermions interact with spin-fluctuations with a phenomenological Action
\begin{eqnarray}
S_{\rm int} = g^2\sum_{{\bf q}, {\bf k} {\bf k'}, i, \alpha, \beta, \gamma, \delta} \sum_{\omega_n} \chi( \vec{q}, i \omega_n) \psi_{{\bf k'-q}, \gamma}^+ {\bf \sigma}_{\gamma, \delta}^i \psi_{{\bf k'}, \delta}\psi_{{\bf k+q}, \alpha}^+ {\bf \sigma}_{\alpha, \beta}^i \psi_{{\bf k}, \beta} + H.C. 
\end{eqnarray}
$\chi$ will be chosen to have dimensions of inverse of energy (after subsuming a factor of $4\mu_B^2$ in its definition). So $g$ is a coupling function of dimension of energy. $g$ for heavy fermions is the exchange energy between the conduction electrons and the $f$- local moments. Its meaning for d-band problems is more ambiguous, and may be best inferred from independent experiments, for example the resistivity above $T_c$.

A suitable phenomenological form for the dynamical spin-fluctuations due to AFM correlations, with which experimental results \cite{stockert1, stockert2} can be fitted,  is
\begin{eqnarray}
      \chi( \vec{q}, i \omega) &=& \frac{\bar{\chi_0} \Gamma_{\rm AFM} }{ \psi_{\vec{q}} + \vert \omega \vert },\\
      \psi_{\vec{q}} &\equiv& \Gamma_{\rm AFM} \left[ (\xi/a)^{-2} + a^2\left(  \vec{q} - \vec{Q} \right)^2 \right].
      \label{chi}
\end{eqnarray}
where $\Gamma_{\rm AFM}$ is the damping rate of the fluctuations, $Q$ is the antiferromagnetic vector.

The correlation length $\xi$ is related to the deviation from the Quantum Critical Point (QCP) by variation in pressure, doping, magnetic field, etc. as well as by temperature. $\Gamma_{\rm AFM}$, ${\bf Q}$ and $\xi$ may all be determined from experiments. The temperature dependence of $\xi$ has been  studied by renormalization group (RG)  \cite{hertz,millis} and by the self-consistent renormalization (SCR) methods \cite{moriya1, moriya2}.
 $\xi^{-2} \propto T^{3/2}$ near AFM QCP in the 3 dimensions and dynamical critical exponent $=2$.
SCR  derives using the same dynamical critical exponent that near the magnetic QCP, $\xi^{-2}(T) \sim \chi_{\vec{Q}} \propto T + \theta$. 
\par
In our calculations in two dimensions $\xi$ will be assumed to be of the form
\begin{eqnarray}
\label{corr2}
      (\xi/a)^{-2}(T) &=& (\xi^{*}/a)^{-2} + \gamma \frac{T}{\Gamma_{\rm AFM}},
	\label{xi}
\end{eqnarray}
where $\xi^{*}$ is the asymptotic $T=0$ value of the correlation length. One of our results is that the temperature dependence of $\xi$ is of insignificant consequence in determining $T_{\rm c}$.
 
The linearized Elisahberg equations give that the kernel for  Cooper pair coupling in the d-wave channel in a square lattice is proportional to the projection of 
 $|g({\bf k,k}')|^2 \chi({\bf k-k}', \omega)$ 
  to $(\cos k_x-\cos k_y) (\cos k_x'-\cos k_y')$. In spin-fluctuations theories,  $|g({\bf k,k}')|^2$ has only a smooth momentum dependence. So, Cooper-pair coupling prominently depends  depends only on (i) the momentum dependence of $\chi({\bf k-k}', \omega)$ determined by the correlation length $\xi$ (ii) the integrated weight in the momentum dependent part and (iii) the energy scale of the momentum dependent fluctuations.  The first is qualitatively obvious from the fact that a q-independent spin - fluctuation contributes zero to the Cooper channel in the d-wave channel. It is not possible to make quantitative statements on these effects without detailed calculations because the results also depend on the nesting in the band-structure near the AFM ${\bf Q}$. We will show that the three ingredients in $\chi({\bf k-k}', \omega)$ are not mutually independent.
  
To gain physical insight, the effect of $\xi$ on the integrated spectral weight may be  discussed before detailed calculations through the the partial sum-rule on $\chi_{\rm AFM}(\vec{Q}, \omega)$, which determines the effective coupling constant for superconductivity:
\begin{align}
\notag	\sum_i \langle S_i^2 \rangle_{\rm AFM} &= \frac{1}{\pi} \sum_{\vec{q}} \int_0^{\omega_c} {\rm d} \omega {\rm Im} \chi ( \vec{q}, \omega)\\
&= \frac{\omega_c }{\pi^2} \bar{\chi^0} \left[ \frac{\pi}{2} - {\rm Tan}^{-1}\left( \frac{\Gamma_{\rm AFM}(\xi/a)^{-2} }{ \omega_c} \right) - \frac{1}{2} \frac{\Gamma_{\rm AFM}( \xi/a)^{-2} }{\omega_c} \log \left[ 1 + \left( \frac{\Gamma (\xi/a)^{-2} }{\omega_c} \right)^{-2} \right]  \right].
	\label{tmsr}
\end{align}
With the assumed
Lorentzian form, it is necessary to introduce an upper cut-off $\omega_c$ in the frequencies $\omega$ up to which the fluctuations extend. Actually,  spin fluctuation are actually quite suppressed for $\omega \sim \Gamma_{\rm AFM} $ and we can simply use $\omega_c \approx \Gamma_{\rm AFM}$ in calculations of Eliashberg equations. It is important to take into account that there are four equivalent AFM-vector for the two dimensional problem in the paramagnetic regime of the model, however strongly fluctuating it may be.
This has been taken into account in the sum-rule by multiplying the measured Im$\chi(\vec{q}, \omega)$ by 4. For $d=3$, the number of equivalent AFM vectors is larger and a correspondingly larger multiplicative factor should be used.

\begin{figure}[ptb]
      \begin{center}
            \includegraphics[width = 0.60\textwidth]{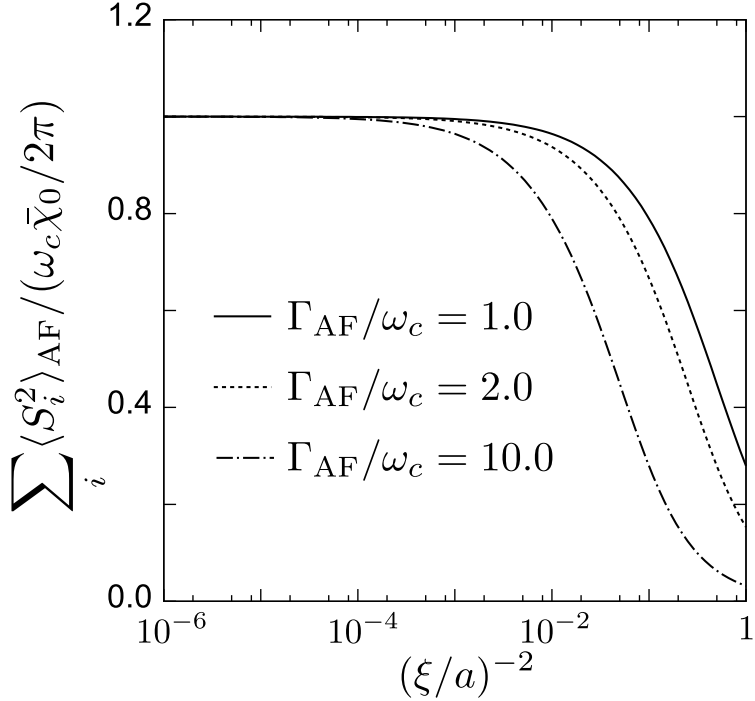}
            \caption{The dependence of the quantity $\displaystyle \sum_i \langle S_i^2 \rangle_{\rm AFM}/ \omega_c (\bar{\chi}_0/2 \pi)$ which is shown in the text to be approximately proportional to the effective coupling constant $\lambda_{eff}$ on the correlation length $\xi/a$ is exhibited for various values of $\Gamma_{\rm AFM}/\omega_c$ shown.}
		\label{fig:00}
	\end{center}
\end{figure}

In the regime of very long correlation lengths, $(\xi/a)^{2} \gg 1$, i.e. close to the quantum-critical point, the sum rule simply gives
\begin{equation}
      \sum_i \langle S_i^2 \rangle_{\rm AFM} \approx \frac{\omega_c }{2\pi} \bar{\chi^0} + O(a/\xi)^2
\end{equation}
$\langle S_i^2 \rangle_{\rm AFM}$ may to a first approximation be estimated from the ordered moment $\langle S \rangle $ in nearby AFM phase but more properly from integration of the relevant momentum and frequency range of the measured fluctuations in absolute units using polarized neutrons.
Fig. \ref{fig:00} shows the $(\xi/a)^{-2}$-dependence of $\displaystyle \sum_i \langle S_i^2 \rangle_{\rm AFM}/ \omega_c (\bar{\chi}_0/2 \pi)$ for $\Gamma_{\rm AFM}/\omega_c$=1.0, 2.0 and 10.0.

Let us now consider the sum-rule in the opposite limit, that the correlation length is small compared to the lattice constant, i.e. the system is very far from the quantum critical point. Then
\begin{equation}
      \sum_i \langle S_i^2 \rangle_{\rm AFM} \approx \frac{\omega_c }{2\pi} \bar{\chi^0} \frac{1}{2}[ (\xi/a)^2+ O(\xi/a)^4] 
\end{equation}

As already shown by ML and further elaborated below, for a given band-structure, the results of the Eliashberg calculations for $T_{\rm c}/\Gamma_{\rm AFM}$ may be parametrized in terms of a dimensionless "bare" coupling constant $\lambda$ and a correlation length $\xi$, 
\begin{equation}
\label{lambda}
\lambda = g^2 N_{\rm F} \bar{\chi}_0 
\end{equation}
$\bar{\chi}_0$ may be determined in terms of  $\langle S_i^2 \rangle_{\rm AFM}$ and therefore (approximately) to the ordered moment through the sum-rule. We may define an effective coupling constant $\lambda_{\rm eff}$ to incorporate the effect of the correlation length. Using that the sum-rule becomes the total moment sum-rule in the limit of infinite correlation length and the maximum possible ordered moment, i.e. that of the AFM insulator (ignoring the zero-point effects), one concludes that in the limit of very  large correlation lengths
\begin{equation}
\label{lambdainfty}
\lambda_{\rm eff} \to \lambda_{\infty} = g^2 N_{\rm F} f^2\langle S_i ^2\rangle_{\rm max} \frac{2\pi}{\omega_c},
\end{equation}
where $f$ is the fraction of the maximum possible ordered moment. From Fig. (\ref{fig:00} and from the detailed calculations presented in the next section, one deduces that the limit for $\lambda_{\infty}$ is reached for $\xi/a \gtrsim 10$, below which there is an exponential fall off of $T_{\rm c}/\Gamma_{\rm AFM}$.
For smaller correlation-lengths, Fig. (\ref{fig:00}) shows that the $\lambda_{\rm eff}$ for $T_{\rm c}$ decreases with decreasing correlation length. 

In the work of ML, $\lambda$ values from about 5 to about 50 are used in the calculations with varying correlation lengths. Actually, one obtains for the considerations of the sum-rules above  that for spin-(1/2) problems, even the coupling constant $\lambda_{\infty}$ is only of $O(1)$, because $g N_{\rm F}$ is of $O(1)$ and so is the upper limit on the ratio $\Gamma_{\rm AFM}/\omega_c$.
An independent estimate of $\lambda_{\rm eff}$ may be obtained from the normal state properties, for example the coefficient of the temperature dependence of the resistivity of non-fermi-liquid form in the quantum critical region. Again only $\lambda$ of $O(1)$ will be found consistent. 

It is also important to note that these BCS type coupling constants do not carry information on  the retardation effects due to the frequency dependence of the interaction; these as well as the effects of inelastic scattering which are particularly important for anisotropic superconductors are properly treated through the numerical solution of the Eliashberg equations. The difference from electron-phonon induced superconductivity where a single parameter $\lambda$ need by introduced \cite{mcmillan} should also be noted. 

It should also be pointed out that in some heavy fermions, the quantum-critical fluctuations do not have the functional form given by  the simple RG or SCR approximations as above, but displays "local criticality" \cite{lqcp} as suggested for the cuprates \cite{mfl}. In this paper, we only consider fluctuations which are well specified by the form given above.

\section{Resistivity in the Quantum-critical Region}

The temperature dependence of the  resistivity near the quantum-critical points has been derived several times \cite{Rosch}. Here, we rederive it paying special attention to the coefficient in front of the anomalous temperature dependent part. An expression of the resistivity $\rho(T)$ in the antiferromagnetic quantum critical region suitable for heavy fermions may be derived with the following formula derived from the Boltzmann equation.
\begin{equation}
      \rho^{-1}(T) = \frac{1}{ 4\pi^3} \frac{e^2 v_{\rm F} }{\hbar} \frac{1}{3} \int \tau_{\vec{k}} dS_{\rm FS},
      \label{aq1}
\end{equation}
where the integration is taken over the fermi-surface, and $v_{\rm F}$ the fermi velocity.
This assumes that the {\it actual} electronic structure near the chemical potential is sufficiently complicated that in the temperature region of interest, vertex corrections which lead to emphasis on large momentum scattering for resistivity are unimportant. In that case the scattering rate which determines the resistivity is the same as the single-particle scattering rate averaged over the fermi-surface. This is true in a multi-sheeted fermi-surface and is suitable for heavy fermions. This is similar to the case of transition metals where the resistivity from electron-phonon scattering at low temperatures is $\propto T^5$ in contrast to the nearly free-electron metals where it is $\propto T^3$. For weakly anisotropic single band scattering, as in the cuprates, the resistivity for large AFM correlation lengths is close to the Fermi-liquid temperature dependence although near the hot spots the scattering rate is nearly $\propto T$ \cite{Hlubina-Rice}.

Equation (\ref{aq1}) can also be expressed as follows.
\begin{equation}
      \rho^{-1} = \frac{ n e^2 }{ m^{*}} \langle \tau_{\vec{k}} \rangle_{\rm FS}.
      \label{aq2}
\end{equation}
where $\langle \cdots \rangle_{\rm FS} \equiv \frac{1}{4 \pi k_{\rm F}^2} \int \cdots dS_{\rm FS}$ means the average over the fermi-surface, $m^{*}$ the renormalized effective mass.\par
Here, $\tau_{\vec{k}}$ can be derived from the imaginary part of the self-energy.
\begin{equation}
      \frac{ \hbar}{ 2 \tau_{\vec{k}} } = - {\rm Im} \Sigma ( \vec{k}, \varepsilon + i\delta)\vert_{\varepsilon \rightarrow 0},
      \label{aq3}
\end{equation}
where the self-energy due to the antiferromagnetic quantum fluctuations is given as follows.
\begin{equation}
      \Sigma ( \vec{p}, i \varepsilon_n)= g^2 k_{\rm B} T \sum_{\omega_m} \sum_{\vec{q}} G( \vec{p}- \vec{q}, i\varepsilon_n - i\omega_m) \chi( \vec{q}, i\omega_m) 
      \label{aq4}
\end{equation}
The result for the resistivity in the limit $\xi/a \to \infty$ is derived in an Appendix A. by explicitly calculating the self-energy given by eq. (\ref{aq4}) is
\begin{equation}
      \rho(\xi/a = \infty) = \lambda \frac{3}{4 e^2} \frac{a \hbar}{ (\varepsilon_{\rm F}/k_{\rm B}) \sqrt{ \Gamma_{\rm AFM}/k_{\rm B} }} T^{3/2}.
      \label{rho}
\end{equation}

\section{Solution of the Linearized Eliashberg Equations}

The superconducting transition temperature is given by the linearized version of the Eliashberg Equations for the normal self-energy $- i \omega_nZ(\theta_{\vec{k}},i \omega_n)$ and the anomalous or pairing self-energy $W( \theta_{\vec k}, i \omega_n )$.
\begin{align}
	\label{leliash1}
	\left[ 1 - Z(\theta_{\vec{k}},i \omega_n) \right] i \omega_n &= - \int_{\rm FS} \frac{ {\rm d}^{d} S_{\vec{p}} }{ (2 \pi)^d v_{\vec{p}} } \pi T \sum_{\Omega_m} i\ {\rm sgn}(\Omega_m) g^2 \chi ( \vec{k} - \vec{p},i \omega_n - i\Omega_m),\\
	\label{leliash2}
	W( \theta_{\vec k}, i \omega_n ) &= - \int_{\rm FS} \frac{ {\rm d}^{d} S_{\vec{p}} }{ (2 \pi)^d v_{\vec{p}} } \pi T \sum_{\Omega_m} \frac{ W( \theta_{\vec p}, i \Omega_m )}{ \vert \Omega_m Z( \theta_{\vec{p}}, i \Omega_m) \vert} g^2 \chi ( \vec{k} - \vec{p},i \omega_n - i\Omega_m).
\end{align}
Here $\omega_n$ are the Matsubara frequencies; $g$ is a momentum-independent coupling matrix element, which has already been defined , $\theta_{\vec{k}}$ is an angle parameterizing the Fermi surface, $N(\theta_{\vec{k}})$ is the density of states at angle $\theta_{\vec{k}}$. The $\vec{p}$-integral is over the Fermi surface, $v_{\vec{p}}= \partial \varepsilon_{\vec{p}}/ \partial \vec{p}$ is the unrenormalized velocity.

\subsection{Results for variation of $T_{\rm c}$ with Parameters in the Models}

Our principal general results for $T_{\rm c}$ on the basis of solution of the linearized Eliashberg equations in terms of $\lambda$ and the parameters in $\chi_{\rm AFM}(\vec{Q}, \omega)$ are given in this section. The numerical evaluation is done by first simplifying the Eliashberg equations (\ref{leliash1}) and (\ref{leliash2}), as far as possible analytically. The final expressions for the numerical evaluation, both for the circular Fermi-surface and the tight-binding Fermi-surfaces are given in Appendix B.\par

Figure. \ref{fig:03a} show the $(\xi^{*}/a)^{-2}$-dependences of $T_{\rm c}/\Gamma_{\rm AFM}$ for the circular fermi-surface on the  bare coupling constant $\lambda$ in the large correlation length limit on the right. For small bare coupling $\lambda$, the latter does have the BCS form while for $\lambda \gtrsim 1$, the dependence is approximately linear. Consistent with earlier discussions \cite{ml1,ml2}, $T_{\rm c}/\Gamma_{\rm AFM}$ shows a very shallow peak at around $(\xi^{*}/a)^{-2}\sim 5 \times 10^{-3}$. 
$T_{\rm c}/\Gamma_{\rm AFM}$ shows a drop-off as the correlation length decreases, while it also shows moderate decreases as the correlation length increases.
ML pointed out that this moderate decrease is caused by the rapid diverges of $Z$ as the correlation length increases. We note also that the quantum-classical crossover correction to the correlation length proportional to the factor $\gamma$ in Eqs. (\ref{corr2}) has a negligible effect on $T_{\rm c}$. This will not be considered in any further calculations.

The principal message from Fig. (\ref{fig:03a}) is that the infinite correlation length result for $T_{\rm c}$ is well obeyed up to $\xi/a \approx 10$ with a very sharp fall off thereafter which will be seen later to be exponential. For large $\xi/a$, no BCS type approximation for $T_{\rm c}$ is valid. The limit of very large correlation length is equivalent to the  effective frequency of fluctuations  $\to 0$, as may be seen from Eq.(\ref{chi}). If we use the McMillan \cite{mcmillan} type approximation, in which  $\lambda_{\rm M} \propto \langle \omega^2 \rangle^{-1}$, the inverse of the average squared frequency of fluctuations, we get a divergent coupling.  Fig. (\ref{fig:03a}) gives a finite limit to $T_{\rm c}/\Gamma_{\rm AFM}$, which depends on the bare coupling constant $\lambda$. One may understand this result from the calculations of Allen and Dynes \cite{Allen-Dynes}, deduced  for s-wave Eliashberg equation, that in the limit of a diverging coupling constant $T_{\rm c} \propto \sqrt{\lambda_{\rm M}} \langle \omega \rangle$, where $\langle \omega \rangle$ may be taken approximately to be the square-root of  $\langle \omega^2 \rangle$. 
-

\begin{figure}[ptb]
      \begin{center}
            \includegraphics[width = 0.75\textwidth]{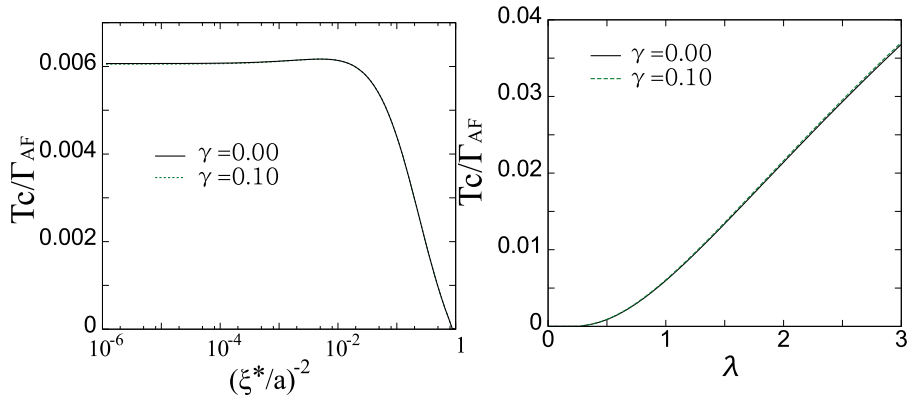}
            \caption{ The transition temperature normalized to $\Gamma_{\rm AFM}$ vs $(\xi^{*}/a)^{-2}$ for a circular fermi-surface and $\lambda=1$ is shown on the left, and
            the transition temperature normalized to $\Gamma_{\rm AFM}$ vs $\lambda$ for a circular fermi-surface and $(\xi^{*}/a)^{-1} = 0$ is shown on the right. }
		\label{fig:03a}
	\end{center}
\end{figure}



\begin{figure}[tb]
	\begin{center}
	      \includegraphics[width = 0.65\textwidth]{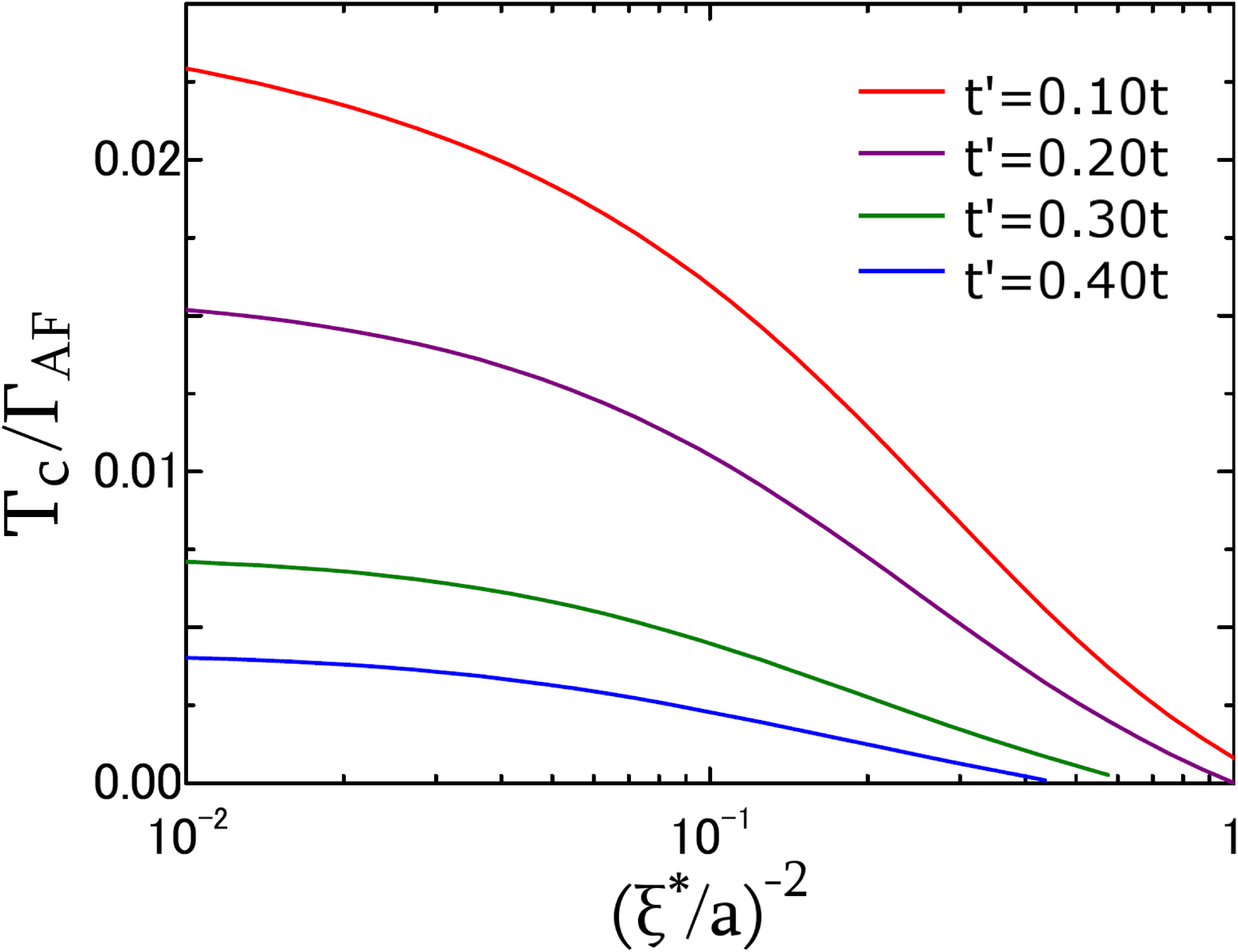}
            \caption{The transition temperature normalized to $\Gamma_{\rm AFM}$ vs $(\xi^{*}/a)^{-2}$ for the four fermi-surfaces shown in Fig. (\ref{fig:01}).}
		\label{fig:04a}
	\end{center}
\end{figure}

Next we show in Fig. (\ref{fig:04a}) the $(\xi^{*}/a)^{-2}$-dependence of $T_{\rm c}/\Gamma_{\rm AFM}$ for the four fermi surface shown in Fig. (\ref{fig:01}) .
For $t'=0.4t$, the worst nesting case, $T_{\rm c}/\Gamma_{\rm AFM}$ is similar to that for the circular fermi-surface.
Improving the nesting condition  increases , $T_{\rm c}/\Gamma_{\rm AFM}$. 

\begin{figure}[tbp]
	\begin{center}
            \includegraphics[width = 0.75\textwidth]{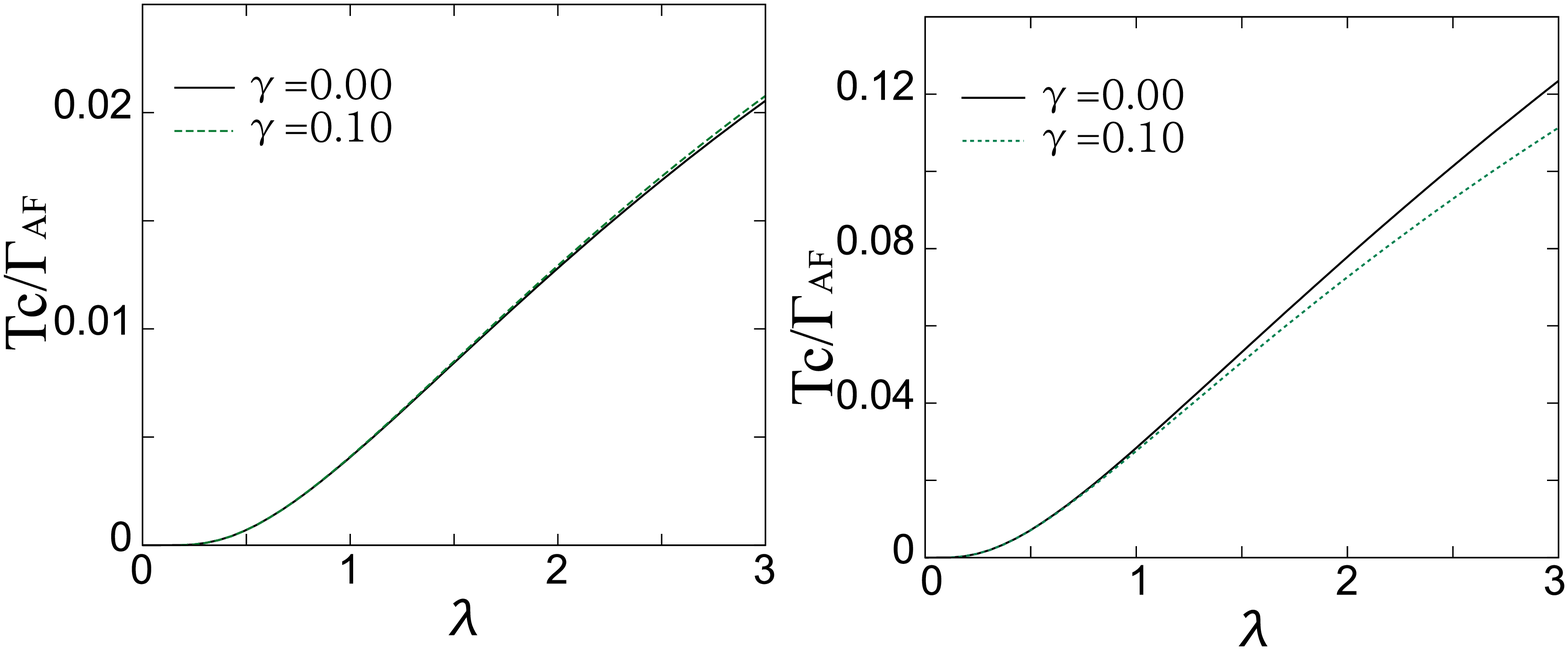}
            \caption{Left: The transition temperature normalized by $\Gamma_{\rm AFM}$ vs $\lambda$ for the worst nested fermi-surface (FS1)  of 
            Fig (\ref{fig:01});  Right: The transition temperature normalized by $\Gamma_{\rm AFM}$ vs $\lambda$ for the best nested fermi-surface (FS4)  of 
            Fig (\ref{fig:01})}. 
		\label{fig:05a}
	\end{center}
\end{figure}

We show in (\ref{fig:05a}) $T_c$ as a function of $\lambda$ for the worst nesting fermi-surface and  the best nested fermi-surface of those in Fig. (\ref{fig:01}). A increase of $O(2)$ in $T_{\rm c}$ for the similar values of $\lambda$ is discerned from the worst to the best nesting conditions.

ML have also presented detailed results for calculations on a 3 d electronic dispersion with the symmetry of a cubic lattice.
They remark that other parameters being the same 2 dispersion gives higher $T_{\rm c}$ than a three dimensional dispersion.
Based on our results for changes in $T_{\rm c}$ in the 2 d problem, we conclude that this is because of the much better nesting that is obtainable in model 2 d systems compared to the 3d systems for a given ${\bf Q}$ which spans the Fermi-surface in some (usually symmetry) direction. In fact, we can place our 2 d-results for the very weakly nested fermi-surface over the ML results for the 3d Fermi-surface and find for other parameters the same that the systematics of the results for $T_{\rm c}$ as well as its value is very similar.

\section{Comparison with Experiments in Heavy Fermions}
\begin{figure}[tb]
	\begin{center}
	      \includegraphics[width = 0.75\textwidth]{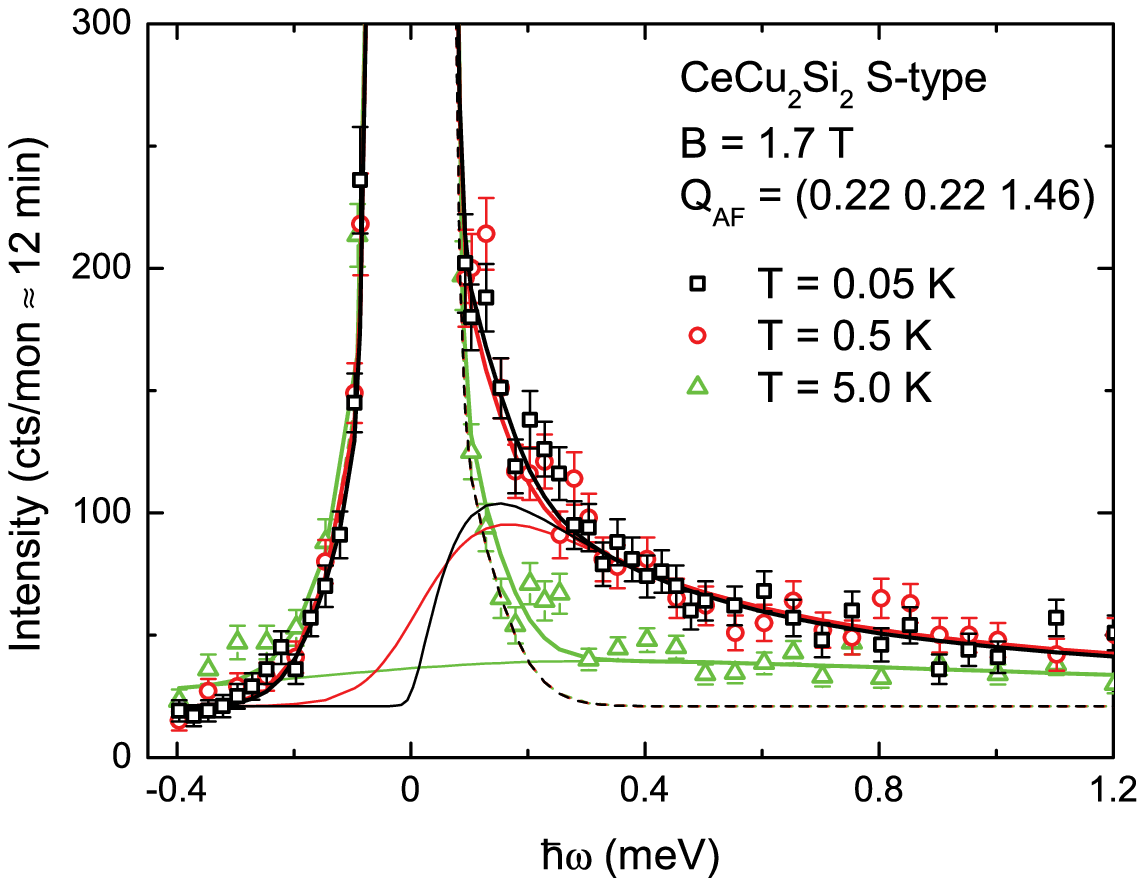}
            \caption{The measured dynamic structure factor at the antiferromagnetic Bragg vector as a function of energy for various temperatures in CeCu$_2$Si$_2$. From Ref. (\onlinecite{arndt}). The variation with ${\bf Q}$ to get the correlation length is also available in (\onlinecite{arndt}) and references therein.}.
		\label{fig:S(q)}
	\end{center}
	\end{figure}

In this section, we compare the estimates of $T_{\rm c}$ from the calculations with the experimental  result in CeCu$_2$Si$_2$ and CeIrIn$_5$. For convenience, we show the measured intensity \cite{arndt} proportional to the dynamic structure factor $S({\bf Q}, \omega)  = \coth (\omega/2T) {\rm Im} \chi({\bf Q}, \omega)$ in Fig. (\ref{fig:S(q)}) for ${\bf Q } $ near ${\bf Q}_{\rm AFM}$.

Although the magnetic fluctuation spectrum found through inelastic scattering in CeCu$_2$Si$_2$ is well represented by the form of Eq. (\ref{chi}), the electronic structure is far more complicated than assumed here or in the 3 d calculations of ML. 
We have seen that $T_{\rm c}$, especially in the limit of large magnetic correlation lengths depends only on gross parameters like $\lambda$ and secondarily on the amount of nesting. The comparison can only be very limited and can only give insight into the orders of magnitudes expected and to the physics involved.
\paragraph{CeCu$_2$Si$_2$}\cite{stockert1,arndt}\\
\noindent
To fit the phenomenological susceptibility to these results, the parameters take the following values: \\
\noindent
	{\it AFM wave vector}: 
		$Q_{\rm AFM}$= (0.22, 0.22, 1.46)\\
	 {\it Ordered Moment in the AFM phase} : 
	        $\langle S \rangle \approx 0.2 \mu_B$/Ce
	 {\it Correlation length:}
		$\xi \simeq$ 25 \AA;\\
   {\it Characteristic energy scales } :\\
		$\Gamma_{\rm AFM} \simeq $ 1.5 [meV];  \\
	{\it Density of States}:	From the measured uniform $(\bf q =0)$ Paramagnetic susceptibility  in the normal state, one deduces the $N(E_{\rm F}) \approx$ (1/7) [meV]$^{-1}/$unit-cell.\\
	{\it AFM spin-fluctuations parameter}:	$\bar{\chi}_0$:  \\
 The experimental results for  $\chi(q,\omega)$  in Ref.(\onlinecite{stockert1}) are parametrized in terms of three quantities $\xi, \chi_0$ and $\Gamma$.  The correlation length $\xi$ in Ref.(\onlinecite{stockert1}) is the same quantity used by us. For clarity we give here the relation of the other two parameters to the parameters used by us. The conversion from the quantity $\chi_0$, which we will call $\chi_{0,S}$ to our $\bar{\chi}^0$ is obtained by equating the integral over all $q,\omega$ of Equation S4 in Ref.(\onlinecite{stockert1}) to the integral of the same physical quantity given in Eq. (\ref{tmsr}). In the limit of $(\xi/a)^2 >> 1$, one gets $\bar{\chi}^0 \approx \chi_{0,S} (a/\xi)^2$. The experimental result is $\chi_{0,S} = 15.64 \mu_B^2/meV$. This then gives $\bar{\chi}^0  \approx 0.4 \mu_B^2/unit-cell/meV$ . \\
The quantity $\Gamma$  is related to $\Gamma_{AFM}$ by $\Gamma_{AFM} = \Gamma (a)^{-2}$.\\ 
{\it Transition Temperature} : 
  $T_{\rm c} \sim$ 0.6[K] \\
CeCu$_2$Si$_2$ has a very anisotropic fermi-surface with very little dispersion along the tetragonal axis. The fermi-surface in the plane is very complicated but we assume that just as in s-wave superconductivity \cite{mcmillan}, $T_{\rm c}$ depends only on the average density of states at the Fermi-surface, supplemented by knowledge of nesting of the fermi-surface near ${\bf Q}_{AFM}$. Among other things, our results below may be taken to be test of this assumption.
\paragraph{CeIrIn$_5$}\cite{kambe,yashima}\\
\noindent
Although long-range magnetic order competing with superconductivity in CeIrIn$_5$ has not been accessed in this stoichiometric compound, there are strong experimental results indicating that the compound lies in the vicinity to an AFM quantum-critical point. The resistivity of this material exhibits a non-Fermi liquid behavior similar to that observed in CeCoIn$_5$, which is known to lie in the vicinity of an AFM quantum-critical point which has been accessed by doping the compound.  Moreover, the nuclear spin relaxation rate of CeIrIn$_5$ is also similar to that of CeCoIn$_5$.
The dynamical susceptibility has been recently deduced by NMR experiments \cite{kambe} in agreement with this conclusion.\\
To fit the susceptibility to these experimental result, the parameters take the following values.\\
	{\it AFM wave vector}: 
		$Q_{\rm AFM}$= (0.5, 0.5, 0.5)\\
 The chosen $Q_{\rm AFM}$ and ordered moment are taken to be that of the related compound CeCoIn$_5$ \cite{stock} \\
	 {\it Ordered Moment in the AFM phase}:
            $\langle S \rangle \approx 0.15 \mu_{\rm B}$\\
          {\it Correlation length}\\
            $(\xi^{*}/a) \simeq 10$ at $T=1$[K]\cite{kambe}
	 {\it Characteristic energy scales of AFM:}
            $\Gamma_{\rm AFM} \simeq$  1.5[meV]; \\
	 {\it Transition Temperature}:
       $T_{\rm c}$ = 0.4[K] 
\ \par
The experimental results show that both CeCu$_2$Si$_2$ and CeIrIn$_5$ lie not far from the asymptotic large correlation length limit and that their $T_{\rm c}/\Gamma_{\rm AFM}$ are both about 0.03. For a circular fermi-surface, and using the measure value of $\xi/a$ in the former, we may refer then simply to Fig. (\ref{fig:03a}) and find that $\lambda \approx 3$ gives the right value of $T_{\rm c}$. For the best nested Fermi-surface, however, a $\lambda \approx 1$ is sufficient as shown in Fig. (\ref{fig:05a}). 

 We may now try to estimate $\lambda$ to see if these values are reasonable.  We do this in two different ways. To utilize the neutron scattering results for this purpose, we need to know $g$ besides the directly measured properties listed above. The renormalizations in the heavy fermion problem are such that near the critical point the AFM  interaction between magnetic moments is of the same order as the heavy fermion bandwidth. Then $g$ is of the order of the effective fermi-energy, i.e $gN_{\rm F}\sim 1$.
Then we may use the experimental values of $N_{\rm F}$ and $\bar{\chi_0}$ deduced from experiments above in Eq.(\ref{lambda}) to get $\lambda \approx 2$. 
 
The above manner of estimation has forced us to guess the value of $g$. We can estimate the value of $\lambda$ much better and independently from the non-fermi-liquid resistivity proportional to $T^{3/2}$ observed in the quantum critical regime of CeCu$_2$Si$_2$, whose coefficient is proportional to $\lambda$. 
The resistivity $\rho$ in the quantum-critical region for $\xi/a \to \infty$  is given by Eq.(\ref{rho}).
Using the values of CeCu$_2$Si$_2$ mentioned above, the resistivity is estimated $\rho = 3.02 \times 10^{-8} \lambda T^{3/2}$ [$\Omega {\rm m}$] from eq. (\ref{rho}).
The non-fermi liquid resistivity observed in CeCu$_2$Si$_2$ takes the form: $\rho(T)/\rho_{\rm 300K}=0.151 + 0.071 T^{3/2}$ \cite{gegenwart} where $\rho_{\rm 300K} \sim 70\mu \Omega {\rm cm}$ \cite{schneider}.
From the comparison of the coefficient of $T^{3/2}$-term in the resistivity between theoretical and experimental results, $\lambda$ is estimated as $\lambda \sim 1.6$. This should be considered an important evidence for the rather obvious  idea that fluctuations that determine the normal state scattering also determine $T_{\rm c}$, and of the consistency of the present calculations. The extent to which the calculations correctly estimate $T_{\rm c}$ may be judged from the fact that in the range of $\lambda$ from the different estimates for it, $T_{\rm c} \propto \lambda$. \\

We comment briefly on an estimation of the condensation energy due to superconductivity and its comparison with the increase in energy of AFM fluctuations on entering superconductivity \cite{stockert1}. The latter has been estimated to be almost a factor of 20 larger than the superconducting condensation energy. The suggestion has been offered that this factor of 20 may be the increase in kinetic energy. In BCS theory for electron-phonon interactions, the absolute magnitude of the change in kinetic and in potential energy are both of the same order as the condensation energy. So, a good reason has to be found for this factor of 20.  We do not have a solution to this enigma.

\section{Summary}

We have presented a solution to the linearized Eliashberg equations using a phenomenological spin-fluctuation spectrum  and simple fermi-surfaces to highlight the important parameters that determine $T_{\rm c}$ for d-wave symmetry. Careful attention has been paid to the partial sum-rule on the q-dependent part of the spin-fluctuation spectra to estimate the effective coupling constant which depends on parameters such as the total partial spectral weight, the correlation length and the upper frequency cut-off of the q-dependent spin-fluctuations. These parameters are not independent and we show their relationship in the simple model studied. With regard to the electronic structure, a knowledge of the average density of states at the fermi-surface is sufficient for determining $T_{\rm c}$ in the s-wave channel \cite{mcmillan}. But for d-wave superconductivity through exchange of well correlated spin-fluctuations, this must be supplemented by a knowledge of nesting. The results for the general solutions are employed for two heavy fermion compounds using their measured spin-fluctuation spectra and other quantities such as specific heat and magnetic susceptibility. Correct estimates for $T_{\rm c}$ to factors of 
$O(2)$ are obtained. Confidence in these results is bolstered by getting the correct observed temperature dependence of the anomalous $T^{3/2}$ resistivity with a coefficient using the same parameters, again correct to factors of $O(2)$. This puts a semi-quantitative backbone to the surmise made long ago that d-wave superconductivity in  such heavy fermions is promoted by large amplitude spin-fluctuations with large correlation lengths such as occur near some AFM quantum critical points.\\

{\it Acknowledgements:} Part of the work by SN was done while visiting University of California, Riverside with the aid of the Global COE program (G10) from the Japan Society for the Promotion of Science. The work of KM is partially supported by a Grand-in-Aid for Scientific Research on Innovative Area ``Heavy Electrons'' (No. 20102008) and a Grant-in-Aid for Specially Promoted Research (No. 20001004) from the Ministry of Education, Culture, Sports, Science and Technology, Japan. The work of CMV is partially supported by NSF under grant DMR-1206298.\\

\noindent
{\bf Appendix A: Derivation of Resistivity Near the Antiferromagnetic Quantum Critical Point}\\

An expression for the resistivity under assumptions suitable for heavy fermions with a multi-sheeted fermi-surface and/or sufficient impurity scattering \cite{Maebashi} is given by eq. (\ref{rho}) in terms of the self-energy function Eq. (\ref{aq4}).
Here, we derive the relation (\ref{rho}) explicitly.
Substituting $\chi( \vec{q}, i \omega_m)$ in the spectral representation into eq. (\ref{aq4}) and carrying out the $\omega_m$-summation, one gets
\begin{equation}
      \Sigma ( \vec{p}, i \varepsilon_n)= - \frac{g^2}{2} \sum_{\vec{q}} \int_{ -\infty}^{\infty} \frac{dx}{ \pi} \frac{ {\rm Im} \chi( \vec{q},x) }{x - i\varepsilon_n + \xi_{\vec{p}-\vec{q}}}  \left(  \tanh \frac{\xi_{\vec{p} -\vec{q} }}{2k_{\rm B}T} + \coth \frac{x}{2k_{\rm B}T} \right).
      \label{aq5}
\end{equation}
Taking the analytic continuation of $\Sigma(\vec{p}, i\varepsilon_n)$, the imaginary is given as
\begin{align}
      \notag      {\rm Im} \Sigma ( \vec{p}, \varepsilon + i\delta) &= - \frac{g^2}{2} \sum_{\vec{q}} \int_{ -\infty}^{\infty} \frac{dx}{ \pi} {\rm Im} \chi( \vec{q},x) \pi \delta(x - \varepsilon + \xi_{\vec{p} -\vec{q}}) \left(  \tanh \frac{\xi_{\vec{p}-\vec{q}}}{2k_{\rm B}T} + \coth \frac{x}{2k_{\rm B}T} \right)\\
      \notag      &=- \frac{g^2}{2} \sum_{\vec{q}} {\rm Im} \chi( \vec{q}, \varepsilon - \xi_{\vec{p}-\vec{q}} )  \left(  \tanh \frac{\xi_{\vec{p}-\vec{q}}}{2k_{\rm B}T} + \coth \frac{\varepsilon - \xi_{\vec{p} -\vec{q}} }{2k_{\rm B}T} \right)\\
      &=- \frac{g^2}{2} \sum_{\vec{q}} \frac{ \bar{\chi_0}\Gamma_{\rm AFM}^{-1} (\varepsilon - \xi_{\vec{p} - \vec{q}} ) }{ [(\xi/a)^{-2} + a^2 (\vec{q} - \vec{Q} )^2 ]^2 + \left( \frac{ \varepsilon - \xi_{\vec{p} - \vec{q}}}{ \Gamma_{\rm AFM}} \right)^2  }  \left(  \tanh \frac{\xi_{\vec{p-q}}}{2k_{\rm B}T} + \coth \frac{\varepsilon - \xi_{\vec{p-q}} }{2k_{\rm B}T} \right).
      \label{aq6}
\end{align}
We now consider the behavior at around the antiferromagnetic quantum critical point, i.e., $ (\xi/a)^{-1} \sim 0$.
In a low temperature region where the non-fermi liquid behavior appears, $\varepsilon \sim 0$ gives the dominant contribution for eq. (\ref{aq6}).
Moreover, using the following relation,
\begin{equation}
      \tanh \frac{x}{2} - \coth \frac{x}{2} = \frac{-2}{\sinh{x}} ,
      \label{aq7}
\end{equation}
eq. (\ref{aq6}) is transformed as
\begin{align}
      \notag      {\rm Im} \Sigma ( \vec{p}, 0 + i\delta) &= - g^2 \bar{\chi_0}\Gamma_{\rm AFM}^{-1} \sum_{\vec{q}} \frac{ \xi_{\vec{p} - \vec{q}} }{ a^4 (\vec{q} - \vec{Q} )^4  + \left( \frac{ \xi_{\vec{p} - \vec{q}}}{ \Gamma_{\rm AFM}} \right)^2  }  \frac{1}{\sinh( \frac{\xi_{\vec{p}-\vec{q}}}{k_{\rm B}T}) }\\ 
      &= - g^2 \bar{\chi_0}\Gamma_{\rm AFM}^{-1} \sum_{\vec{q}^{'}} \frac{ \xi_{\vec{p} - \vec{Q} - \vec{q}^{'} } }{ a^4 \vec{q}^{'4} + \left( \frac{ \xi_{\vec{p} - \vec{Q} - \vec{q}^{'} }}{ \Gamma_{\rm AFM}} \right)^2  }  \frac{1}{\sinh( \frac{\xi_{\vec{p}- \vec{Q} - \vec{q}^{'} }}{k_{\rm B}T}) }. 
      \label{aq8}
\end{align}
Next, we consider the $\vec{q}$-integration in eq. (\ref{aq8}).
Because the denominator in eq. (\ref{aq8}) has $\vec{q}^{'4}$ term, $q^{'} \sim 0$ gives the dominant contribution in the $\vec{q}^{'}$-integration.
Therefore, one gets
\begin{align}
      {\rm Im} \Sigma ( \vec{p}, 0 + i\delta) &= - g^2 \bar{\chi_0}\Gamma_{\rm AFM}^{-1} \frac{1}{2 \pi^2} \int_0^{q_c} d q^{'} \frac{ \xi_{\vec{p} - \vec{Q}} }{ a^4 q^{'4} + \left( \frac{ \xi_{\vec{p} - \vec{Q} }}{ \Gamma_{\rm AFM}} \right)^2  }  \frac{1}{\sinh( \frac{\xi_{\vec{p}- \vec{Q} }}{k_{\rm B}T}) }. 
      \label{aq9}
\end{align}
Since the integrated function in eq. (\ref{aq9}) rapidly decays as $q^{'}$ increases, we take $q_c$ as $\infty$ and obtain following result by easy calculation.
\begin{equation}
      {\rm Im} \Sigma ( \vec{p}, 0 + i\delta) = - g^2 \bar{\chi_0}\Gamma_{\rm AFM}^{-1/2} \frac{1}{8 \sqrt{2} \pi a^3 } \frac{\vert \xi_{\vec{p} - \vec{Q}} \vert^{1/2} }{\sinh( \frac{\xi_{\vec{p}- \vec{Q} }}{k_{\rm B}T}) }. 
      \label{aq10}
\end{equation}\par
Substituting eqs. (\ref{aq10}) and (\ref{aq3}) into eq. (\ref{aq2}), $\rho$ is given as
\begin{equation}
      \rho \simeq \lambda \frac{a \hbar}{2 \sqrt{2} \Gamma_{\rm AFM}^{1/2} e^2} \langle \frac{ \vert \xi_{\vec{p} - \vec{Q}} \vert^{1/2} }{ \sinh \frac{\xi_{\vec{p} - \vec{Q}}}{k_{\rm B} T} } \rangle_{\rm FS},
      \label{aq11}
\end{equation}
where we use $na^3 \sim 1$ and $N_{\rm F} = m^{*} k_{\rm F}/(2 \pi^2 \hbar^2)$.\par
\begin{figure}[tb]
	\begin{center}
		\includegraphics[width = 0.5\textwidth]{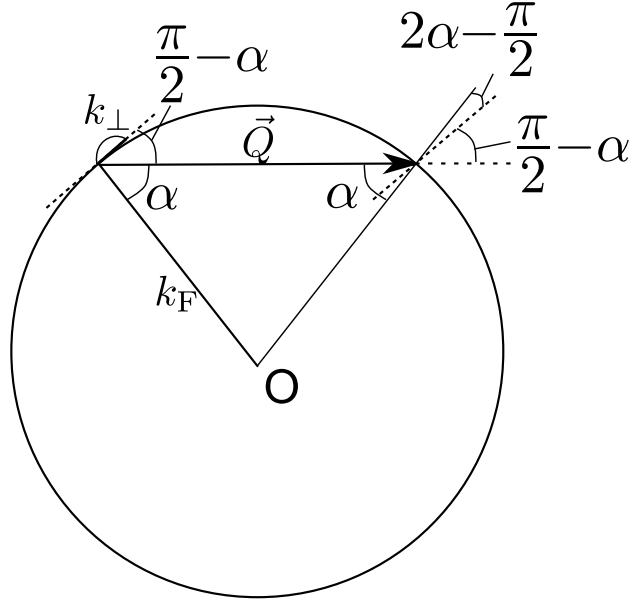}
	\end{center}
      \caption{The cross-sectional circular fermi-surface and the wave vector $\vec{p}$ which satisfies the relation $\vert \vec{p} \vert = \vert \vec{p} - \vec{Q} \vert = k_{\rm F} $.}
	\label{fig:11}
\end{figure}

Here, we estimate the average over the fermi-surface in eq. (\ref{aq11}) assuming that the fermi-surface is spherical.
\begin{equation}
      \langle \frac{ \vert \xi_{\vec{p} - \vec{Q}} \vert}{ \sinh \frac{\xi_{\vec{p} - \vec{Q}}}{k_{\rm B} T} } \rangle_{\rm FS} = \frac{1}{ 4 \pi k_{\rm F}^2 } \int \frac{ \vert \xi_{\vec{p} - \vec{Q}} \vert^{1/2} }{ \sinh \frac{\xi_{\vec{p} - \vec{Q}}}{k_{\rm B} T} } d S_{\rm FS}.
      \label{aq12}
\end{equation}
The dominant contribution in eq. (\ref{aq12}) comes from ``hot'' line where the relation $\vert \vec{p} \vert = \vert \vec{p} - \vec{Q} \vert = k_{\rm F} $ is satisfied as shown in Fig. \ref{fig:11}
Assuming that the dispersion near the fermi-surface is given by linear dispersion, we obtain $\xi_{\vec{p} - \vec{Q}} \simeq v_{\rm F} k_{\perp} \cos( 2 \sigma - \pi/2) = k_{\rm F} \sin 2 \alpha$, where $k_{\perp} $ is the deviation from the ``hot'' line.
For one ``hot'' spot, the integration is estimated as 
\begin{equation}
      \frac{1}{ 4 \pi k_{\rm F}^2 } k_{\rm F} \sin \alpha \int_{-k_c}^{k_c} d k_{\perp} \frac{ \sqrt{ \vert v_{\rm F} k_{\perp} \sin 2 \alpha \vert}  }{ \sinh \frac{ v_{\rm F} k_{\perp} \sin 2 \alpha }{k_{\rm B} T} }. 
      \label{aq13}
\end{equation}
Changing the integration variable as $x \equiv v_{\rm F} k_{\perp} \sin 2\alpha/( k_{\rm B} T)$, eq. (\ref{aq13}) is transformed as
\begin{equation}
      \frac{ (k_{\rm B} T)^{3/2}  }{ 8 \pi v_{\rm F} k_{\rm F} \cos \alpha }  \int_{0}^{ \frac{v_{\rm F} k_c \sin 2 \alpha}{ k_{\rm B}T }} d x \frac{ \sqrt{x}}{ \sinh x}.
      \label{aq14}
\end{equation}
Now, we take the upper limit of the integration as $\infty$ because we consider the low temperature region, and eq. (\ref{aq14}) can be calculated as 
\begin{equation}
      \frac{ (k_{\rm B} T)^{3/2}  }{ 8 \pi v_{\rm F} k_{\rm F} \cos \alpha }  \frac{ 2\sqrt{2} - 1}{ \sqrt{2}} \zeta \left( \frac{3}{2} \right)  \Gamma \left(  \frac{3}{2} \right)
      \simeq \frac{ 3 (k_{\rm B} T)^{3/2}  }{ 4 \pi v_{\rm F} k_{\rm F} \cos \alpha }  
      \label{aq15}
\end{equation}\par
Since such a ``hot'' point makes two rings whose total length is equal to $4 \pi$ in the sphere fermi-surface, eq. (\ref{aq12}) is given by 
\begin{equation}
      \langle \frac{ \vert \xi_{\vec{p} - \vec{Q}} \vert}{ \sinh \frac{\xi_{\vec{p} - \vec{Q}}}{k_{\rm B} T} } \rangle_{\rm FS} = \frac{ 3 (k_{\rm B} T)^{3/2}  }{ 2 \varepsilon_{\rm F} \cos \alpha }.  
      \label{aq16}
\end{equation}\par
Substituting eq. (\ref{aq16}) into eq. (\ref{aq11}), we obtain
\begin{equation}
      \rho \simeq \lambda \frac{3 a \hbar}{4 \sqrt{2} \cos \alpha } \frac{1}{ (\varepsilon_{\rm F}/k_{\rm B}) (\Gamma_{\rm AFM}/k_{\rm B}) ^{1/2} e^2}  T^{\frac{3}{2}}.
      \label{aq17}
\end{equation}
In this calculation, the $Q$-vector is given by $2k_{\rm F} \sin \alpha$.
The $Q$-vector of the CeCu$_2$Si$_2$ is observed as (0.215, 0.215, 0.1458) giving $\vert \vec{Q} \vert = 1.49 a/\pi$. 
Therefore, $\alpha$ is estimated as $ \alpha \sim \pi/4$, and we obtain the result used for the estimation of $\lambda$:
\begin{equation}
      \rho \simeq \lambda \frac{3 }{4e^2 } \frac{ a \hbar}{ (\varepsilon_{\rm F}/k_{\rm B}) (\Gamma_{\rm AFM}/k_{\rm B}) ^{1/2} }  T^{\frac{3}{2}}.
      \label{aq18}
\end{equation}

\noindent
{\bf Appendix B: Final Expressions for Evaluation of $T_{\rm c}$}\\

\noindent
{\it Circular Fermi-surface}\\

For a circular Fermi-surface, it is possible to do the momentum integrals in the Eliashberg equations (\ref{leliash1}) and (\ref{leliash2}) analytically so that only a diagonalization in discrete frequency space needs to be done numerically. The final expressions used for numerical evaluation for the normal and the anomalous self-energy are:

\begin{align}
      \label{f1}
      Z(\theta_{\vec{k}},i \omega_n) &=  1 + \frac{ \lambda}{ \omega_n/(\pi T)} \sum_{\Omega_m}  \frac{{\rm sgn}(\Omega_m)}{\sqrt{ \alpha^2 - \beta^2}}  \\
      \label{f2}
      W_{2}(i \omega_n) &= \pi T \sum_{\Omega_m} K( \omega_n, \Omega_m) \frac{ W_{2} (i\Omega_m)}{ \vert \Omega_m \vert },\\ 
      \label{f3}
	K(\omega_n, \Omega_m) &= - \lambda \int_{0}^{2\pi} \frac{ {\rm d} \theta_{\vec p}}{2 \pi}\frac{ \cos 2 \theta_{\vec{p}} \cos2x }{ \vert Z( \theta_{\vec{p}}, \Omega_m) \vert} \frac{ A - \sqrt{ A^2 - B^2 - C^2} }{ (B^2 + C^2) \sqrt{ A^2 - B^2 - C^2}}, 
\end{align}
where 
\begin{eqnarray}
      \alpha &=&  \frac{\vert \omega_n - \Omega_m \vert}{ \Gamma_{\rm AFM} } + (\xi/a)^{-2} + a^2 ( \vert \vec{k} \vert^2 + \vert \vec{p} \vert^2 + \vert \vec{Q} \vert^2) - 2 a^2 \vert \vec{k} \vert \vert \vec{Q} \vert \cos(\theta_{\vec{k}} - \theta_{\vec{Q}} ), \\
      \beta &=& 2 a^2 \vert \vec{p} \vert \sqrt{ \vert \vec{Q} \vert^2 + \vert \vec{k} \vert^2 - 2 \vert \vec{Q} \vert \vert \vec{k} \vert \cos ( \theta_{\vec{k}} - \theta_{\vec{Q}} ) },\\ 
      A &=& \frac{\vert \omega_n - \Omega_m \vert}{a^2 \Gamma_{\rm AFM} } + (\xi/a)^{-2} + a^2 ( \vert \vec{k} \vert^2 + \vert \vec{p} \vert^2 + \vert \vec{Q} \vert^2) + 2a^2  \vert \vec{p} \vert \vert \vec{Q} \vert \cos(\theta_{\vec{p}} - \theta_{\vec{Q}} ), \\
      B^2 + C^2 &=& 4 a^4  \vert \vec{k} \vert^2 \left[ \vert \vec{p} \vert^2 + \vert \vec{Q} \vert^2 + 2\vert \vec{p} \vert \vert \vec{Q} \vert \cos ( \theta_{\vec{p}} - \theta_{\vec{Q}} )\right],\\
      x &=& \tan^{-1} \left[ \frac{ \vert \vec{Q} \vert \sin \theta_{\vec{Q}} + \vert \vec{p} \vert \sin \theta_{\vec{p}}} { \vert \vec{Q} \vert \cos \theta_{\vec{Q}} + \vert \vec{p} \vert \cos \theta_{\vec{p}}} \right].
	\label{f4}\\
\end{eqnarray}\par 

\noindent
{\it  Tight-Binding Fermi-surfaces}\\

With tight binding approximation, only some simplifications in the momentum integrals in the Eliashberg equations (\ref{leliash1}) and (\ref{leliash2}) can be done analytically. The final expressions used in this paper for numerical evaluation are
\begin{align}
	\label{f5}
      Z(\theta_{\vec{k}},i \omega_n) &=  1 + \frac{1}{ \omega_n/(\pi T)} \sum_{\Omega_m} {\rm sgn}(\Omega_m) \int_{FS} \frac{ {\rm d}^2 S_{\vec{p}} }{(2 \pi)^2 v_{\vec{p}}} \frac{ \lambda }{ \frac{\vert \omega_n - \Omega_m \vert}{\Gamma_{\rm AFM}}  + (\xi/a)^{-2} + a^2 ( \vec{k} - \vec{p} - \vec{Q} )^2 }, \\
      W_{2}(i \omega_n) &= \pi T \sum_{\Omega_m} K( \omega_n, \Omega_m) \frac{ W_{2} (i\Omega_m)}{ \vert \Omega_m \vert },\\ 
      K( \omega_n,\Omega_m ) &\equiv - 2\lambda \int_{FS} \frac{ {\rm d}^2 S_{\vec{k}} }{(2 \pi)^2 v_{\vec{p}}} \int_{FS} \frac{ {\rm d}^2 S_{\vec{p}} }{(2 \pi)^2 v_{\vec{p}}} \frac{ 1 }{ \vert Z(\theta_{\vec{p}} ,i\Omega_m ) \vert }\frac{ \big(\cos (k_{Fx}a) -   \cos (k_{Fy}a)\big)\big(\cos (p_{Fx}a) -   \cos (p_{Fy}a)\big)}{ \vert \omega_n - \Omega_m \vert/\Gamma_{\rm AFM} + (\xi/a)^{-2} + a^2( \vec{k} - \vec{p} - \vec{Q} )^2 }.
	\label{f6}
\end{align}

For both circular and tight binding Fermi-surfaces, the best numerical strategy to evaluate $T_{\rm c}$ is to cast Eqs. (\ref{f3}) and (\ref{f6}) in the form of an eigenvalue equation for the eigenvector $W/ \vert \omega_n \vert $
\begin{align}
      \sum_{\Omega_m} \left[ K(\omega_n,\Omega_m) - \frac{ \vert \omega_n \vert}{ \pi T} \delta_{n,m} \right] \left[ \frac{W(i \Omega_m) }{ \vert \Omega_m \vert }  \right] = 0.
	\label{a9}
\end{align}
\par

It should be noted that the Matrix of eq. (\ref{a9}) is not Hermitian because $K(\omega_n, \Omega_n)$ includes the renormalization factor $Z(\omega_n, \theta_{\vec{k}})$.
If the angle dependence of $Z$ can be neglected, we can define $K$ in a form which does not include $Z$, and we obtain the eigenvalue equation with a Hermitian Matrix for the eigenvector $W/ \vert \omega_n Z(\omega_n) \vert$.
In s-wave superconductor case, such a situation, namely angle-independent self-energy appears.
However, in the d-wave case, $Z(i \omega_n, \theta_{\vec{k}})$ strongly depends on $\theta_{\vec{k}}$.
On including $Z$ in the kernel $K$, the latter is no longer symmetric for the frequency exchange, $\omega_n$ and $\Omega_m$.\par
At high temperatures the eigenvalues of eq. (\ref{a9}) are close to the negative odd integers.
As the temperature decreases, the largest eigenvalue increases and crosses zero at transition temperature $T = T_{\rm c}$.\\
\par

\end{document}